# An All-Optical Metro Network Architecture and QoS-Aware Wavelength Allocation Study for Converged Fixed, Mobile, and Edge Computing Multi-Granular Traffic

David Georgantas[1], Zhaoyang Liu[2], Georgios Drainakis[1,3], Bitao Pan[2], Adonis Bogris[4], Peristera Baziana[1*]
[1]Department of Informatics and Telecommunications, University of Thessaly, Greece
[2]State Key Lab of Information Photonics and Optical Communications, Beijing University of Posts and Telecommunications, P.R. China
[3]School of Electrical and Computer Engineering, National Technical University of Athens, Greece
[4]Department of Informatics and Computer Engineering, University of West Attica, Greece
dageorgantas@uth.gr, zhaoyang.liu@bupt.edu.cn, gdrainakis@mail.ntua.gr, bitao.pan@bupt.edu.cn, abogris@uniwa.gr, baziana@uth.gr*

*Abstract*—In this paper, we introduce an all-optical metro network architecture, called MOON, to serve converged multi-granular traffic from fixed, mobile, and edge computing services. Since traffic is characterized by high dynamicity and diverse access requirements, MOON uses network slicing to provide quality of service (QoS) aware wavelength allocation to fulfill the various applications traffic demands. MOON incorporates hybrid optical switching (HOS) combining optical circuit switching (OCS) and optical time slotted switching (OTS) capabilities that appropriately maps different traffic types to them. Specifically, the OCS network slice explicitly serves aggregated traffic of long duration and high volume, while OTS network slice serves short bursty traffic. In order to provide flexibility, separate sets of wavelengths are used for OCS and OTS traffic service, both within a metro-access network (MAN) (intra-MAN) and between different MANs (inter-MAN). We extensively study the required number of wavelengths to efficiently serve OCS and OTS traffic for intra- and inter-MAN communication scenarios, taking into account their specific traffic access requirements in an effort to optimize wavelengths utilization. In our study, we assume non-blocking OCS communication for immediate access; therefore the number of required OCS wavelengths depends only on the number of nodes, while the number of required OTS wavelengths to obtain a desired QoS and latency level is independent from the number for OCS wavelengths. Simulation results show that within an OTS intra-MAN we achieve end-to-end (E2E) latency in sub-milliseconds scale, suitable for dynamic bursty traffic, while it is an decreasing function of the number of used OTS wavelengths.

## I. INTRODUCTION

The gradual prevalence of novel bandwidth-demanding applications, such as real-time communications, artificial intelligence (AI) and machine learning (ML), internet-of-things (IoT) etc., has caused significant traffic increase in networks at the access and metro scale, as well as at the edge. At the metro scale, optical networking technologies are exploited to serve the relative traffic explosion due to the high bandwidth provided. In this framework, optical metro networks (OMNs) need to provide connectivity among radio access nodes, edge computing nodes, and residential users. Such a heterogeneous network requires high levels of dynamicity and flexibility, and low and deterministic latency [1]. An all-optical network with routable optical connections in metro and access segments is considered a promising solution to avoid massive optical-electrical-optical (OEO) processing and further achieve reduced latency and power efficiency. Recently, there have been research efforts on optical circuit switching (OCS) based all-optical networks [2-3]. However, dynamic traffic with sub-milliseconds latency requirements produced by novel network applications (e.g. industry control etc. [4]) in metro-access networks (MAN) demands fast and highly reliable data delivery. Current OCS networks have long (tens of milliseconds) reconfiguration time and high wavelength consumption; hence they are not suitable to deploy communication resources for dynamic traffic. To address the issues of wavelength limitation and slow response to traffic requests in OCS based networks, efforts have been made on optical time-slot switching (OTS) technologies [5-6]. Despite the potential advantages of the OTS, challenges arise when deploying it in OMNs that encompass both MANs and metro core networks (MCNs). Nevertheless, OTS can work well in a small scale network with a simple topology and a local supervisory channel control. When it comes to OMNs with many nodes and hierarchical topology, the supervisory channel fails to handle all the OTS operations due to the lack of coordination between hierarchies. Therefore, both OCS and OTS based technologies have failed to deploy end-to-end (E2E) all-optical networking in OMNs while meeting the requirements of dynamic and high demanding access applications.

In order to effectively serve the dynamic traffic in optical MANs and MCNs, this work introduces MOON [7] and extensively explores its wavelength utilization enhancement. MOON is an all-optical metro network with hybrid optical switching (HOS) capability designed for efficient wavelength usage and rapid response to dynamic traffic requests. In MOON, multi-granular E2E all-optical links can be established throughout the entire network, with OTS for wavelength efficiency and small granular bursty traffic delivery, and OCS for carrying long duration and high volume aggregated traffic. In addition, MOON achieves all-optical networking in OMNs using one network architecture to fit all types of access traffic [7]. The key innovation and contribution of this work is that MOON integrates OCS and OTS, allowing for all-optical networking in OMNs with efficient and on-demand wavelength usage. This enables the generation of multi-granular optical channels to serve multi-granular client traffic, resulting in increased utilization of optical channels and rapid response to bursty traffic. This paper explores the performance of the proposed OTS and OCS network slices in terms of average throughput, E2E latency and dropping rate, for diverse MAN nodes populations and traffic loads. Moreover, this work explores through exhaustive simulations the required numbers of wavelengths used for OTS and OCS to provide guaranteed QoS to the differentiated traffic types. Since for non-blocking OCS communication the number of required OCS wavelengths depends only on the number of nodes, it does not affect the number of required OTS wavelengths to obtain a desired QoS level. Simulation results show that within an intra-MAN the OTS and OCS E2E latency is in sub-milliseconds scale, which makes our proposal suitable for both dynamic bursty OTS traffic, and OCS traffic of long duration and volume.

This paper is organized as follows: Section II presents the MOON architecture. Section III gives the OTS and OCS slice performance evaluation in the intra- and inter-MAN domain. Finally, Section IV concludes our work.

## II. NETWORK MODEL AND ASSUMPTIONS

The access types of MOON network generally include edge computing nodes, fixed optical access nodes, and mobile radio access networks (RAN), as illustrated in Fig. 1(a). Client traffic destined to an intra-MAN or inter-MANs both can be carried by either OCS or OTS. QoS requirements are used to determine which optical network slice should be chosen to serve the traffic.

In MAN segment, MOON has the capability of hybrid and cooperative OTS and OCS over different sets of wavelengths, which are called OTS and OCS intra-MAN wavelengths respectively. With OTS, MOON is able to forward client traffic using connection-less time-slotted wavelengths, thus fast responding to the traffic requests. The OTS wavelengths can be reused at each time slot by all the MAN nodes, reducing wavelength consumption. For the intra-MAN communication, the HOS nodes use a number $W$ of OTS wavelengths, each of which consists of 240 times slots around the MAN ring. Each time slot lasts for 2.1 µs [7]. The intra-MAN ring propagation delay is 504 µs, corresponding to a 100.8 km ring perimeter. We assume that each intra-MAN OTS wavelength data rate is $R_{OTS}$ 25 Gbps. Each HOS is equipped with $W$ optical transceivers one for each of the $W$ OTS intra-MAN wavelengths. Moreover, each HOS keeps a single electrical buffer with size 64 KB that stores the outgoing to the intra-MAN OTS frames.

With OCS, MOON is able to provide E2E optical connections to the client traffic with long duration and high reliability requirements. For the intra-MAN OCS communication, the HOS nodes use a number $N$ of OCS wavelengths, each @$R_{OCS}$=100 Gbps. Each HOS is equipped with $N$ optical transceivers, one for each of the $N$ OCS intra-MAN wavelengths. If we assume that the number of the HOS nodes is $M$, each HOS keeps $M$-1 electrical buffers, each with size 25 KB that stores the outgoing to the intra-MAN OCS frames destined to each of the possible destination HOS nodes. For non-blocking two-way OCS communication, a single OCS wavelength has been explicitly assigned between any given pair of HOS nodes. This means that for non-blocking OCS intra-MAN connectivity, the required number $L$ of lightpaths is $L$= $M\times(M$-1), while the required number $N$ of OCS wavelengths is $N$=$L$/2 for reuse purposes. At any MAN configuration, both the required numbers $L$ of lightpaths and $N$ of OCS wavelengths depend solely on the number $M$ of HOS nodes.

In the MCN segment, reconfigurable optical add-drop multiplexers (ROADMs)/optical cross-connects (OXCs) are responsible for wavelength-level OCS networking, using a set of $K$ OCS inter-MAN wavelengths. If we assume that there is a number of $B$ ROADMs in the MCN segment, each ROADM keeps $B$-1 electrical buffers of 25 KB size to store the outgoing to the MCN OCS frames. The MCN ring propagation delay is assumed to be 5040 µs, corresponding to a 1008 km MCN ring perimeter. Furthermore, the OTS traffic can be optically groomed into a single wavelength which is called OTS aggregation wavelength to be routed by the MCN to other MANs enabling efficient all-optical E2E communication. For the OTS aggregation communication, each HOS is equipped with one optical transceiver running @100 Gbps and with one electrical buffer with size 16KB dedicated to store the outgoing frames destined to another MANs.

Both intra-MAN and inter-MAN communication are orchestrated by a central controller. The network central controller is responsible for making the HOS work properly, by instructing the HOS MAN nodes which wavelengths can be used in OTS. At each node, a local controller receives instructions from the central controller in each 21 μs and directs the wavelength selective switch (WSS) accordingly. Moreover, the local controller is also responsible for receiving and analyzing the supervisory channel and for configuring the OTS switches. Figure 1(b) shows an overview of a MOON network with HOS deployed [7]. The ROADM based MCN nodes are organized in a mesh topology and some MCN nodes are connected to a HOS-based MAN that is organized as a ring or horseshoe topology. The ROADM node can configure the HOS based MANs in different topologies. In case a wavelength channel is used for intra-MAN HOS networking, then the MCN node is configured to make a ring topology for the MAN nodes, as shown in the upper part of Fig. 1(c). Here, $\lambda_1$ and $\lambda_2$ are used for OTS networking in a ring topology, where the continuous optical channels are periodically divided into multiple time slots that are used as sub-wavelength data channels for OTS networking. In addition, the out-of-band supervisory channel is also being configured in the ring topology similar to the data channels for synchronous network control. With the time-slotted optical switching, any two MAN nodes can establish communication paths in a few microseconds, thus at least one wavelength can make an all-connected OTS network.

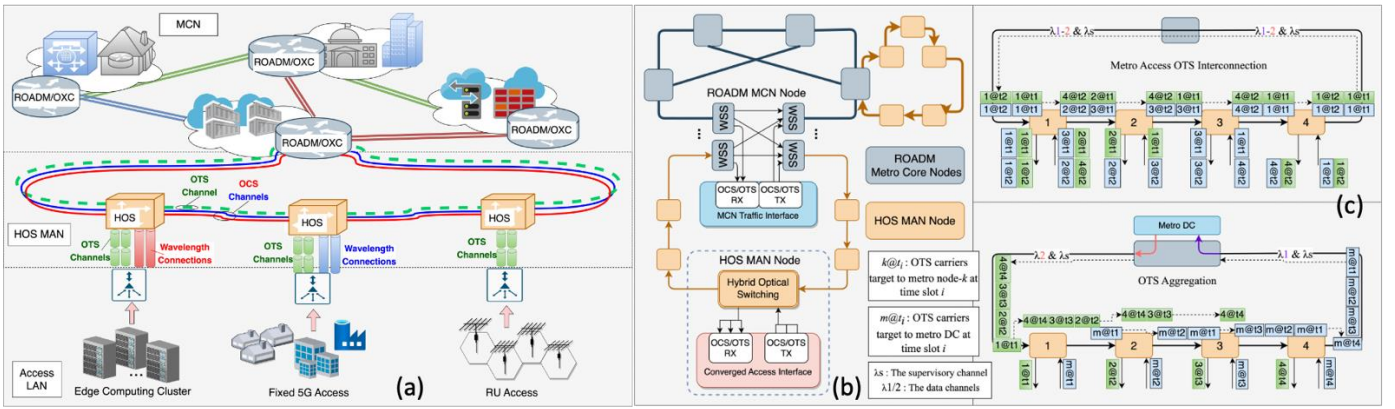

Fig. 1. MOON architecture: (a) MAN and MCN segments, (b) HOS operation, (c) Upper: Intra-MAN OTS wavelength, Lower: MAN-MCN OTS wavelength.

MOON also supports all-optical traffic grooming and distribution between MAN and MCN for OTS inter-MAN communication. Multiple OTS slots can be aggregated as a unit to be routed and forwarded together by a MAN over the OTS aggregation wavelength to another MAN. In this case, the MCN node configures the MAN network in a horseshoe topology as shown in the lower part of Fig. 1(c). Here, $\lambda_1$ is used for traffic grooming from the sending MAN nodes to the MCN node, and $\lambda_2$ is used for traffic distribution from the MCN node to the receiving MAN nodes. Both wavelengths are time slotted, while each slot over $\lambda_1$ can have a different source and each slot over $\lambda_2$ can have a different destination.

For both OTS and OCS traffic used for MOON evaluation, the inter-arrival times of the generated packets follow exponential distribution. Each generated packet is first segmented into 257-bit data blocks, where the first bit is extracted and mapped to the overhead of an optical service unit (OSU) frame [8], while the remaining 256 bits form the payload. The OSU frame is a transport structure designed for fine-granularity service mapping in optical networks, with a total size of 192 bytes, consisting of a 185-byte payload and 7 bytes of overhead [8]. Since a single OSU frame cannot fully accommodate an integer number of 256-bit blocks, a multi-frame approach is used, grouping 32 OSU frames together according to destination and traffic slice (OTS or OCS) to ensure complete transport of the data blocks. The generation time of a frame is defined as the generation time of the last data block of the last OSU frame that enters the multi-frame structure.

III. PERFORMANCE EVALUATION

To evaluate the performance of MOON network slicing, we ran several simulations using a developed Python-based simulator on a machine equipped with an Intel Core i7-1065G7 CPU @1.30 GHz, 12 GB of RAM, and a 64-bit Windows operating system. Our simulator emulates the following different MOON network slices: OTS intra-MAN, OCS intra-MAN, OCS MCN and OTS aggregation. The OCS MCN slice results are not presented in a separate subsection, as they differ from the OCS intra-MAN results only in terms of propagation delay (MCN ring perimeter is greater than the intra-MAN ring perimeter) and are therefore included solely within the overall performance subsection.

Our simulator derives results for each network slice for the performance metrics of: average throughput, average queueing delay (the average time a frame spends in an electrical buffer until its transmission), average total delay (the average sum of queueing and propagation delay) and the average dropping rate (the average rate of dropped frames due to buffers overflow). All simulations were conducted over a total duration of 2.52 ms, corresponding to 1200 simulation events.

A. OTS Intra-MAN Slice

Fig. 2(a) presents the average throughput (left vertical axis) and the average dropping rate (right vertical axis) versus the average offered load for the intra-MAN OTS communication, for $W$=4 OTS wavelengths while the number $M$ of HOS nodes varies, $M$=4, 6, 8, 12. For the same $W$ and $M$ configurations, Fig. 2(b) presents the relative values for the average queuing (left axis) and total delay (right axis) versus the average offered load.

For comprehension purposes, we first study the proposed OTS intra-MAN slice performance for $W$=4 and $M$=4. As Fig. 2(a) shows with the blue solid and dashed lines, for offered load up to 60 Gbps the OTS intra-MAN can effectively serve all the incoming OTS traffic without drops, while the average throughput is almost equal to the offered load. In this load range, the queuing delay is lower than 16.5 μs, as Fig. 2(b) presents. For loads higher than 60 Gbps, OTS intra-MAN gradually reaches saturation. Especially for loads higher than 80 Gbps throughput decreases and dropping rate increases, as Fig. 2(a) shows. This behavior is due the fact that as load increases in this high offered load range, the load is shared among the buffers of only a few HOS nodes ($M$=4) resulting in high buffers overflow

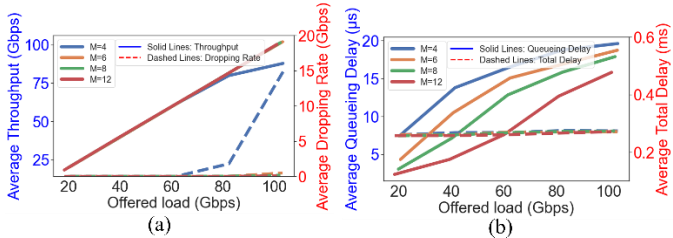

Fig. 2. OTS intra-MAN for $M$=4, 6, 8, 12 and $W$=4: (a) Average Throughput (left axis) and Average Dropping Rate (right axis) vs. Offered Load, (b) Average Queuing Delay (left axis) and Average Total Delay (right axis) vs. Offered Load

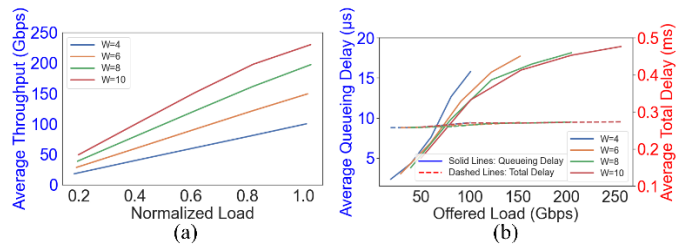

Fig. 3. OTS intra-MAN for $W$=4, 6, 8, 10 and $M$=12: (a) Average Throughput vs. Normalized Load, (b) Average Queuing Delay (left axis) and Average Total Delay (right axis) vs. Offered Load

and consequently high queuing delay and dropping rate, as Fig. 2(b) and 2(a) respectively illustrate. For example for load= 60 Gbps, 80 Gbps and 100 Gbps, throughput is 59.38 Gbps, 78.87 Gbps, 85.82 Gbps respectively, dropping rate is 0, 1.79 Gbps, 14.82 Gbps respectively, and queuing delay is 16.45 µs, 18.76 µs, 19.79 µs respectively.

In Fig. 2(a) and 2(b), we investigate OTS intra-MAN slice performance for a varying number $M$ of HOS nodes varies. As $M$ increases for the same offered load, higher bandwidth utilization is achieved, as Fig. 2(a) depicts. For example, for a load of 80 Gbps, throughput is 78.87 Gbps for $M$=4, 79.64 Gbps for $M$=6, 79.66 Gbps for $M$=8 and 79.67 Gbps for $M$=12. Moreover, as Fig 2(a) also presents, the dropping rate is a decreasing function of $M$ for the same load conditions, while queuing delay also deceases for the same load as $M$ increases (Fig. 2(b)). For example for load 80 Gbps, dropping rate is 1.79 Gbps for $M$=4, and 0 Gbps for $M$=6, 8, 12 respectively, and queuing delay is 18.76 µs for $M$=4, 16.87 µs for $M$=6, 15.83 µs for $M$=8 and 12.65 µs for $M$=12 respectively. This is because as $M$ increases the offered load is distributed among higher number of HOS nodes, resulting in less buffer overflows and lower buffer congestion. It is worth mentioning that queueing delay is lower than 20 µs and total delay is lower than 0.252 ms even under nominal load (100 Gbps) for all HOS nodes configurations, even for low $M$ value ($M$=4), as it can be observed in Fig. 2(b). This means that the proposed HOS intra-MAN architecture can efficiently serve OTS intra-MAN traffic of high variability and time-sensitivity providing sub-milliseconds E2E latency, as required.

Fig. 3(a) studies the average throughput versus the normalized load for $M$=12 HOS nodes while the number $W$ of OTS wavelengths varies $W$=4, 6, 8, 10. Fig. 3(b) illustrates the relative values for the average queueing delay (left axis) and the average total delay (right axis) versus the offered load. As we observe in Fig. 3(a) for any load value, throughput is an increasing function of $W$. This is because as $W$ increases, higher probability of successful transmissions is provided, thus total throughput is increased. For example, for normalized load 0.8, the average throughput is 79.23 Gbps for $W$=4, 118.84 Gbps for $W$=6, 157.68 Gbps for $W$=8 and 193.28 Gbps for $W$=10. Consequently for the same load conditions, as $W$ increases both the queuing and total delay decrease, since the higher OTS intra-MAN capacity can serve faster the incoming frames. This is observed, for load 100 Gbps where the queuing delay is 15.79 µs for $W$=4, 13 µs for $W$=6, 12.12 µs for $W$=8, and 11.91 µs for $W$=10. It is worth mentioning that for all configurations the total delay is in the sub-milliseconds E2E latency scale, totally conforming to the requirements.

### B. OCS Intra-MAN Slice

Fig. 4(a) presents the average throughput (left axis) and the average dropping rate (right axis) versus the normalized load for $M$=4, 6, 8, 10, 12 HOS nodes configuration. Moreover, Fig. 4(b) shows the relative values for the average queueing (left axis) and total (right axis) versus the normalized load. For each $M$ value, the nominal MAN capacity equals to: $L \times R_{OCS} = M \times (M-1) \times R_{OCS}$.

We first analyze the OCS intra-MAN performance of one of the proposed configurations, specifically for $M$=12 ($N$=66). As Fig. 4(a) shows with the purple solid and dashed lines, for normalized load up to 0.8, the OCS intra-MAN can effectively serve almost all offered traffic, which become actual throughput while the dropping rate is almost zero. For normalized loads higher than 0.8, throughput slope reduction and dropping rate slope increase are noticed, due to gradual saturation. For example, the throughput is 10.54 Tbps for normalized load=0.8 and it slightly increases to 12.64 Tbps for load=1.0, while the dropping rate is 0.02 Tbps for load=0.8 and it abruptly increases to 0.54 Tbps for load=1.0. This is because for loads higher than 0.8, the queueing delay abruptly increases due to buffers overflow, as illustrated in Fig 4(b). For example, the queueing delay is 0.27 µs for load=0.8 while it abruptly increases to 0.63 µs for load=1.0.

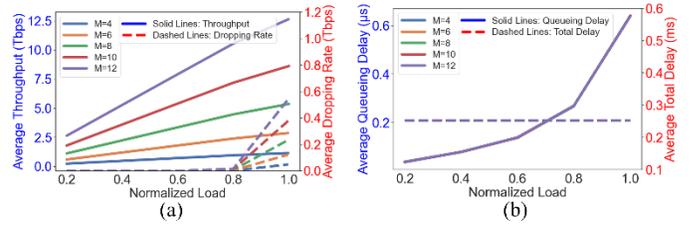

Fig. 4. OCS intra-MAN for $M$=4, 6, 8, 10, 12 and $N$=6, 15, 28, 45, 66 respectively: (a) Average Throughput (left axis) and Average Dropping Rate (right axis) vs. Normalized Load, (b) Average Queuing Delay (left axis) and Average Total Delay (right axis) vs. Normalized Load

It is evident that for non-blocking access, as the number $M$ of HOS increases, the required number $N$ of OCS wavelengths increases too, providing higher transmission probability. This fact causes an increase of the average throughput, as Fig. 4(a) depicts for $M$ variation. Indeed, for the same load conditions, throughput is an increasing function of $M$. For example, as Fig. 4(a) shows for normalized load 0.8, the throughput value is 0.958 Tbps for $M$=4, 2.395 Tbps for $M$=6, 4.472 Tbps for $M$=8, 7.188 Tbps for $M$=10 and 10.54 Tbps for $M$=12. As Fig. 4(a) also presents, for the same load the dropping rate is an increasing function of $M$. This is due to the fact that as $M$ increases the actual offered load increases too, while the dropping probability increases. For example, for normalized load 0.8 the actual offered load is 0.96 Tbps for $M$=4, 2.4 Tbps for $M$=6, 4.48 Tbps for $M$=8, 7.2 Tbps for $M$=10 and 10.56 Tbps for $M$=12, while the relative dropping rate is 1.69 Gbps for $M$=4, 4.23 Gbps for $M$=6, 7.64 Gbps for $M$=8, 11.3 Gbps for $M$=10 and 16.1 Gbps for $M$=12. In other words, for all configurations even under nominal load, bandwidth utilization is high enough while it reaches approximately 95.7%. Notably, for all $M$ values, for any given load the average queueing delay remains constant in sub-microseconds scale. This behavior is due to the fact that for all configurations the buffers utilization remains almost constant for any load value. For example, for all $M$ values, the queueing delay is 0.27 µs for load=0.8 and 0.64 µs for load=1.0. Finally, as Fig. 4(b) shows, the total delay remains in sub-milliseconds scale for all loads and all configurations, as required.

### C. OTS Aggregation Network Slice

For the OTS communication between two different MANs through the MCN, in our simulator we assume that the distance between the two MANs is half of the total MCN ring length (504 km), while the propagation delay between them over the MCN is 2520 µs.

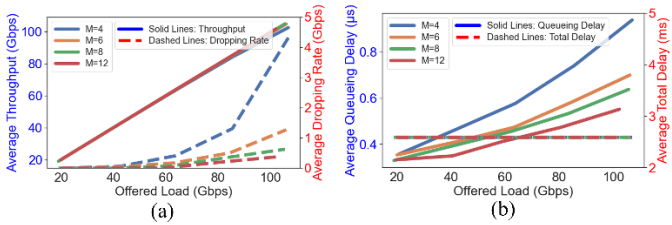

Fig. 5. OTS aggregation for $M$=4, 6, 8, 12 and a single aggregation wavelength: (a) Average Throughput (left axis) and Average Dropping Rate (right axis) vs. Offered Load, (b) Average Queuing Delay (left axis) and Average Total Delay (right axis) vs. Offered Load

Fig. 5(a) presents the average throughput (left axis) and dropping rate (right axis) versus the average offered load for $M$= 4, 6, 8, 12 HOS nodes in each intra-MAN ring. Furthermore, Fig. 5(b) presents the relative values for average queueing (left axis) and total delay (right axis) versus the average offered load.

For comprehensive study, we first explore the OTS aggregation performance for the configuration which employs $M$=4 HOS nodes in each intra-MAN. As Fig. 5(a) shows with the blue solid line, for loads lower than 40 Gbps the OTS aggregated traffic is effectively served without drops, resulting in throughput that is almost equal to the offered load. On the contrary for loads higher than 40 Gbps, the OTS aggregation slice gradually reaches saturation as observed by the slight throughput slope reduction in Fig. 5(a). This trend is also reflected by the dashed blue line that indicates the abrupt increase of the average dropping rate slope for loads higher than 40 Gbps. For example, the dropping rate is 1.08 Gbps for load 80 Gbps while it increases to 3.35 Gbps for load 100 Gbps. This is because at this load range, buffers overflow gradually increases and queuing delay increases too, as Fig. 5(b) presents. For example, queuing delay is 0.7 μs for load 80 Gbps while it increases to 0.87 μs for load 100 Gbps. It is worth mentioning that the total delay across all loads for any configuration is less than 3 ms. This value of only a few ms is slightly higher than the required sub-milliseconds scale, while it is strongly determined by the propagation delay across the connected MANs and the MCN. As such, a configuration of a shorter distance would result in OTS aggregation E2E latency performance at the required sub-milliseconds scale.

The variation of number $M$ of HOS nodes in each MAN significantly affect the OTS aggregation slice performance. Indeed, as $M$ increases the HOS nodes buffers experience fewer buffer overflows since the offered load is shared among more HOS nodes and consequently among more buffers. This is the reason why the dropping rate is a decreasing function of $M$ for loads higher than 40 Gbps. For example, for normalized load 0.8, the dropping rate is 1.08 Gbps for $M$=4, 0.44 Gbps for $M$=6, 0.3 Gbps for $M$=8, 0.19 Gbps for $M$=12, while the relative queuing delay is 0.69 μs for $M$=4, 0.56 μs for $M$=6, 0.52 μs for $M$=8, 0.47 μs for $M$=12. For all configurations and loads, the total delay is less than 3 ms. This means that for other possible configurations of shorter distance among the two MANs through the MCN, it would reach the required sub-milliseconds scale.

### D. Overall Performance

In this subsection, we combinedly study the E2E performance of the different MOON slices. In our simulator, the MOON configuration assumed includes two ring MANs, each with $M$=12 HOS nodes, and a ring MCN comprising $B$=8 MCN nodes. In specific, we combinedly study: a) the OTS intra-MAN communication within each MAN (OTS intra-MAN 1 and OTS intra-MAN 2) over a set of $W$ wavelengths, b) the OCS intra-MAN communication within each MAN (OCS intra-MAN 1 and OCS intra-MAN 2) over a set of $N$=66 OCS wavelengths, c) the OCS communication within the MCN (OCS MCN) over a set of $K$ OCS wavelengths, and d) the OTS aggregated communication between the two MANs (OTS aggregation) over a single OTS aggregation wavelength. The OCS communication within the MCN ring is performed similarly to the OCS intra-MAN communication, over a separate set of $K$ wavelengths where each runs @100 Gbps. The number $K$ only depends on the number $B$=8 of MCN nodes for non-blocking reasons, similarly to the OCS intra-MAN non-blocking operation, i.e. $K$=28. Each MCN node keeps one buffer of 25 KB per any other MCN node for OCS MCN communication. The configuration of the other networks (OTS and OCS intra-MAN and OTS aggregation) is similar to that given in the previous subsections.

Fig. 6 studies the overall performance of MOON (average throughput and queueing delay) for the different network slices as well as in total. Fig. 6(a) presents the average total throughput versus the normalized load for different number $W$ of OTS intra-MAN wavelengths, $W$=4, 6, 8, 10. As illustrated, for any given load there is a gain in total throughput as $W$ increases. For example, for normalized load 0.8, the total throughput is 25.8 Tbps for $W$=4, 25.88 Tbps for $W$=6, 25.96 Tbps for $W$=8, 26.03 Tbps for $W$=10. It is remarkable that this result totally conforms to the relative results of Section III.A since the number $W$ variation only affects the OTS intra-MANs throughput and not the other network slices throughput. In addition, it is worth mentioning that for $W$=10 (that corresponds to a total nominal load of 32.25 Tbps), the total throughput is 31.2 Tbps, which equals to 96.7% total bandwidth utilization. Overall bandwidth utilization minimum value is 95.7% across all configurations (even under nominal load) and reaches a max value of 95.81%.

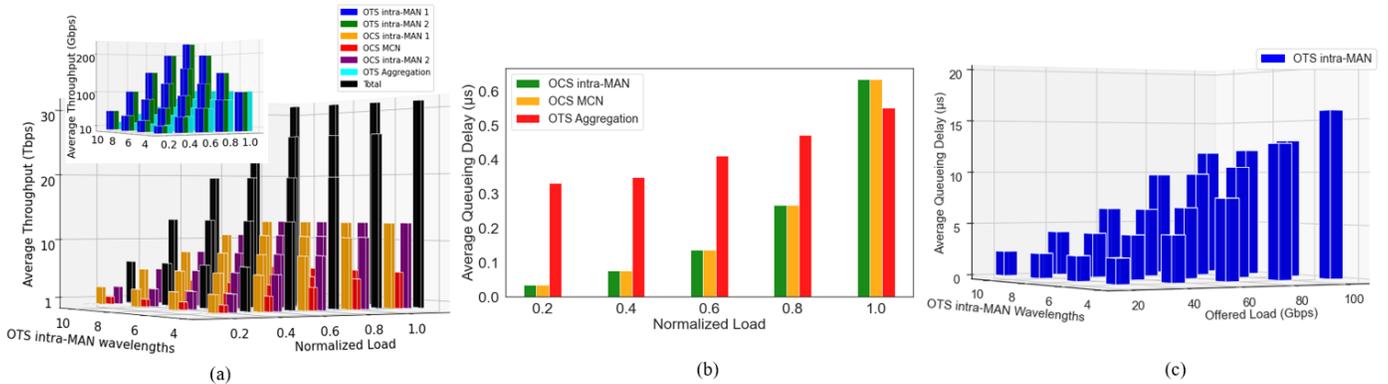

Fig. 6. Total Performance for OTS intra-MAN 1 & OTS intra-MAN 2 ($M$=12, $W$=4, 6, 8, 10), OCS intra-MAN 1 & OCS intra-MAN 2 ($M$=12, $N$=66), OCS MCN ($B$=8, $K$=28), OTS aggregation between MAN 1 and MAN 2: (a) Average Throughput vs. Normalized Load for diverse $W$ (all slices), (b) Average Queuing Delay vs. Normalized Load (OCS intra-MAN, OTS aggregation, OCS MCN), (b) Average Queuing Delay vs. Normalized Load for diverse $W$ (OTS intra-MAN).

Fig. 6(b) illustrates the average queueing delay versus the normalized load for each one of the OCS intra-MANs, for the OCS MCN and for the OTS aggregation slice. All network slices experience an average queueing delay lower than 0.7 μs, even under the highest load of each network slice. Despite keeping different number of nodes, the OCS MCN and OCS intra-MAN slices exhibit nearly identical queueing delay across all loads. This means that for each network slice the buffers are able to almost immediately serve the incoming OCS traffic with low dropping probability. Moreover, Fig. 6(c) presents the average queueing delay versus the offered load for the OTS intra-MAN slice for different number $W$ of OTS intra-MAN wavelengths, $W$=4, 6, 8, 10. Since the nominal load is different in each configuration (100 Gbps for $W$=4, 150 Gbps for $W$=6, 200 Gbps for $W$=8, and 250 Gbps for $W$=10), we choose to compare them under common loads, i.e. up to 100 Gbps. The results reveal that queueing delay is a decreasing function of $W$, as also proved in Section III.A. Representatively, queueing delay experiences 25% reduction when increasing $W$ from 4 to 10 wavelengths.

As presented in Fig. 6(b) and 6(c), for all MOON slices configurations the queueing delay remains at microseconds scale. This means that all buffers across MOON can efficiently store the incoming OTS/OSC traffic with few drops. Notably, for any MOON slice the total E2E delay strongly depends on the propagation delay in each network slice, i.e. on each slice dimensions (the distance between any possible communicating nodes). As demonstrated in Fig. 2, 3 and 4, in the intra-MAN domain both OTS and OCS traffic experience total E2E latency of sub-milliseconds for all configurations, as required. On the other hand, for OTS aggregation and OCS MCN communication the total E2E latency strongly depends on the MCN ring diameter, while it critically increases with the MCN ring diameter increase due to the added propagation delay. Concluding, in order to keep the total E2E latency in sub-millisecond scale as required for novel applications for both OTS and OCS traffic in intra-MAN and MCN domain, the MCN dimensions should be properly determined due to the added propagation delay. Moreover, the number of OTS intra-MAN wavelengths are determined by the required total latency for OTS traffic in the intra-MAN domain, while the number of OCS wavelengths in both intra-MAN and MCN domain is determined by the number of nodes in each network for non-blocking.

## IV. Conclusion

This paper presents MOON, an all-optical metro network architecture designed to combinedly serve converged multi-granular traffic with diverse QoS requirements in terms of E2E latency, coming from different access networks (fixed, mobile etc.), as well as from edge computing services. Across MOON, hybrid optical switching is used to provide differentiated E2E latency delivery based on both OCS and OTS techniques. Therefore, different network slices are assumed: the OTS one serves bursty traffic of small volume, while OCS one serves large volume traffic of long duration. A QoS-aware wavelength allocation study is provided that explores the optimal number of OTS and OCS wavelengths in both intra- and inter-MAN scale, in order to guarantee the desired level of QoS delivery for all traffic types. Especially, it is shown that for OCS communication the required number of wavelengths for non-blocking depends only on the number of nodes, while for OTS communication the E2E latency provided is a decreasing function of the number of OTS wavelengths. Extensive simulations show that even for low number of OTS wavelengths, the E2E latency of dynamic OTS traffic is in the sub-milliseconds scale for any intra-MAN configuration assumed, and in a few (<3) milliseconds scale for inter-MAN communication of long distance (and consequently high propagation delay) between the connected MANs. As a future work, the proposed wavelength allocation study could be further extended counting in different traffic slices within OTS and OCS traffic in order to incorporate traffic diversity in QoS delivery demands, similar to those of novel bandwidth-demanding applications.